\documentstyle[aps,prl,epsf]{revtex} 
\begin{document}
\draft
\tighten
\twocolumn[\hsize\textwidth\columnwidth\hsize\csname
@twocolumnfalse\endcsname
\title{Exchange in silicon-based quantum computer architectures}
\author{Belita Koiller$^{1,2}$,
Xuedong Hu$^1$, and S. Das Sarma$^1$}
\address{$^1$Department of Physics, University of Maryland, College
Park, MD 20742-4111}
\address{$^2$ Instituto de F{\'\i}sica, Universidade Federal do 
Rio de Janeiro, 21945, Rio de Janeiro, Brazil}
\date{\today}
\maketitle
\begin{abstract}
The silicon-based quantum computer proposal has been one of the
actively pursued ideas during the past three years.  Here we 
calculate the donor electron exchange in silicon and germanium, and
demonstrate an atomic-scale challenge for quantum 
computing in Si (and Ge), as the six (four) conduction band minima
in Si (Ge) lead to inter-valley electronic interference,
generating strong oscillations in the exchange splitting of
two-donor two-electron states.  Donor positioning with atomic scale
precision within the unit cell thus becomes a decisive factor in
determining the strength of the exchange coupling---a fundamental
ingredient for two-qubit operations in a silicon-based quantum
computer.  
\end{abstract}
\vspace*{-0.2in}
\vskip2pc]
\narrowtext

Following the seminal proposal \cite{Kane} by Kane there has been a
great deal of activity \cite{Kane1,Vrijen,Clark} in efforts to
develop a silicon-based quantum computer (QC) architecture.  The
basic ideas of the Kane proposal are simple and attractive: to use
donor nuclear spins as quantum bits (qubits), and to utilize
the vast infrastructure and technology associated with the Si
industry to fabricate precisely controlled Si nanostructures, where
exchange effects between electrons and nuclei in neighboring donor
impurities (e.g. $^{31}$P in Si) could serve as the two-qubit gates,
similar to the electron-spin-based QC proposal by Loss and DiVincenzo
\cite{LD}.  The motivation for a Si quantum computer is obvious: 
Once the basic one-qubit and two-qubit operations have been
demonstrated using donor impurities in Si nanostructures, computer
chip fabrication technology associated with the existing and dominant
Si industry will easily enable the scale-up of information processing
involving large number of donor nuclear spin qubits.  Indeed, one of
the formidable stumbling blocks in developing working quantum
computer hardwares has been the scale-up problem, as the demonstrated
qubits in trapped ion and liquid state NMR techniques are not readily
scalable in any significant manner \cite{Reviews}.  

A great deal of experimental work is currently being aimed at
developing suitable qubits in Si nanostructures with precisely
introduced dopant impurities, using both a ``top-down'' approach with
ion-implantation, and a ``bottom-up'' approach with MBE growth and
scanning tunneling microscopy \cite{Clark}.  In the Si QC model 
\cite{Kane,Kane1}, 
donor electrons act as shuttles between different
nuclear spins.  For two-qubit operations, which are required for a
universal QC, both electron-electron exchange and
electron-nucleus hyperfine interaction need to be precisely
controlled.  These are unquestionably formidable experimental
problems.  In the original proposal, Kane used the Herring-Flicker
exchange formula \cite{HF} for two hydrogenic centers to obtain an
order of magnitude estimate of the electron exchange among donors in
Si \cite{Kane}.  However, as he also pointed out, donor
exchange in Si is not hydrogenic.   

In this Letter we show that exchange effects in proposed donor
nuclei based Si QC architectures are actually very subtle due to
quantum interference effects inherent in the complicated Si band
structure.  In particular, special care, going far beyond what is
being currently attempted in the fabrication of Si QC, will be
required in controlling the surface gates crucial to QC operations. 
We also provide results for Ge which seems to have certain advantages
over Si as a candidate for dopant exchange based quantum computation.
Unfortunately, Ge is not such an attractive practical
alternative since there is no well-established Ge technology to take
advantage of.

The study of shallow impurities in Si and Ge is a quite
mature field \cite{Kohn,KL}.  
However, complications arising from the anisotropy of the
electron effective mass and from interference among the Bloch
wavefunctions at the degenerate conduction-band edges were never
fully explored in detail.  Both effects were discussed by
Andres et al \cite{Andres} statistically, in a study of magnetic
susceptibility of Si:P system performed in the context of localization
and magnetic phase transitions. 
Silicon conduction band has six minima located along the 
$<$100$>$ directions, at about 85\% between
the center ($\Gamma$ point) and the boundary (X points) of the
Brillouin zone (BZ): ${\bf k}_{\pm z} = 0.85 (0,0,\pm 2\pi/{\rm a})$,
and the equivalent $x,y$ directions.  The interference between these
valleys causes fast oscillations (on the scale of the inter-atomic
spacing, quite different from the slowly varying exchange interaction
in the spin-based quantum dot QC model \cite{LD,BLD,HD,MultiE}) in the
exchange interaction.  Such oscillations cannot be accounted for by a
simple calculation using hydrogenic centers.  Here we perform a
Heitler-London calculation of the exchange coupling \cite{BLD} between
two substitutional donors in bulk Si, incorporating the effects of
both the anisotropic effective mass and the valley interference. 
Within the envelope function approach, the ground single valence donor
electron state (with $A_1$ symmetry) is written as a symmetric
superposition of the six conduction band minima \cite{Kohn}: 
\begin{equation}
\psi ({\bf r}) = \frac{1}{\sqrt{6}} \sum_{\mu}^6 F_{\mu} ({\bf r})
\phi_\mu(\bf r)\,,
\label{eq:sim}
\end{equation}
where $\phi_\mu({\bf r}) = u_{\mu}({\bf r})e^{i {\bf k}_{\mu}\cdot 
{\bf r}}$ are Bloch wavefunctions.
Here ${\bf k}_{\mu}$ refer to the six conduction band minima, and
$F_{\mu} ({\bf r})$ are the corresponding envelope functions.
We use the Kohn-Luttinger variational form for these anisotropic 
envelopes \cite{Kohn,KL}, e.g.,
\begin{equation}
F_{\pm z} ({\bf r}) = \frac{1}{\sqrt{\pi a^2 b}} \ 
e^{-\sqrt{(x^2+y^2)/a^2 + z^2/b^2}} \,.
\end{equation}
To determine the 
effective Bohr radii $a$ and $b$, we have used the most 
recently measured Si and Ge parameters \cite{madelung}
(especially longitudinal and transverse effective masses) to perform a
variational calculation \cite{KL}.  The calculated wavefunction
widths, as shown in Table~\ref{table1}, are quite close to 
those obtained over 40 years ago \cite{Kohn}. The large difference in
transverse mass has only minimal effect on our results. 

To calculate the exchange splitting between the ground singlet and
triplet states for an impurity pair in Si, we use an approximate
form of the Heitler-London approach \cite{Andres}, which leads to
\begin{eqnarray}
J({\bf R}) & = & \int d^3{\bf r}_1 d^3{\bf r}_2 \psi^* ({\bf r}_1)
\psi^* ({\bf r}_2 - {\bf R}) \frac{e^2}{\epsilon |{\bf r}_1 - {\bf
r}_2|} \nonumber \\
& & \times \psi ({\bf r}_1 - {\bf R}) \psi ({\bf r}_2) 
\label{eq:general}\\
& = & 
\frac{1}{36} \sum_{\mu, \nu} 
\left[ \sum_{{\bf K},{\bf K'}} |c_{\bf K}^{\mu}|^2 |c_{\bf
K'}^{\nu}|^2  e^{i({\bf K}-{\bf K'})\cdot {\bf R}} \right] \nonumber
\\ 
& & \times j_{\mu \nu} ({\bf R})
\cos ({\bf k}_{\mu}\cdot {\bf R}) \cos ({\bf k}_{\nu}\cdot {\bf R})
\,,
\label{eq:exch}
\end{eqnarray}
where ${\bf R}$ is the relative position of the impurity nuclei pair, 
and
\begin{eqnarray}
j_{\mu \nu} ({\bf R}) & = & \int d^3{\bf r}_1 d^3{\bf r}_2 F_{\mu}^*
({\bf r}_1) F_{\nu}^* ({\bf r}_2 - {\bf R}) \frac{e^2}{\epsilon |{\bf
r}_1 - {\bf r}_2|} \nonumber \\
& & \times F_{\mu} ({\bf r}_1 - {\bf R}) F_{\nu} ({\bf r}_2)
\,. 
\label{eq:envelope}
\end{eqnarray}
The second summation in (\ref{eq:exch}) comes from the reciprocal
lattice expansion 
of the periodic part of the Bloch function, $u_{\mu}({\bf r}) = 
\sum_{\bf K} c_{\bf K}^{\mu} 
e^{i {\bf K}\cdot {\bf r}}$, and is identically 1 when ${\bf R}$ 
is an fcc crystal lattice vector.
Integrals in Eq.~(\ref{eq:general}) with rapidly oscillating integrands 
are neglected in Eq.~(\ref{eq:exch}).  For $\mu=\nu$, we calculate
the integrals in (\ref{eq:envelope}) by replacing the denominator
$|{\bf r}_1 - {\bf r}_2|$ by its value for the line along
\begin{table}[h]
\caption[]{Experimental values for the lattice parameter, 
dielectric constant and electron effective masses 
(in units of the free electron mass) for Si and Ge
\cite{madelung}. 
The last two columns give the calculated values of the effective 
Bohr radii.}  
\begin{center}
\begin{tabular}{ccccccc}  
& a (\AA) & $\epsilon$ & $m_\|$ & $m_{\bot}$ & $a$ (\AA) 
& $b$ (\AA) \\  \hline
Si & 5.43 & 12.1 & 0.916 & 0.191 & 25.09 & 14.43 \\ 
Ge & 5.657 & 16 & 1.58 & 0.082 & 64.21 & 22.83 \\ 
\end{tabular}
\end{center}
\label{table1}
\end{table}
\noindent the impurities \cite{Andres}, and assume $j_{\mu \nu} =
\sqrt{j_{\mu \mu} j_{\nu \nu}}$ when $\mu \neq \nu$.
All are excellent approximations for large separations 
($|{\bf R}| \gg a,\,b \gg {\rm a}$).  

Germanium conduction band has four minima along the $<$111$>$ 
directions, at the zone boundary L points. 
Straightforward changes in (\ref{eq:sim})
lead to expressions for the exchange coupling in Ge similar to
(\ref{eq:exch}), with $\mu$ and $\nu$ along the $<$111$>$ directions. 
Because the conduction band minima are located at the zone boundary, 
the oscillation in the exchange coupling should display 
a simpler behavior than in Si when both donors are on the
same fcc sublattice, thus making Ge an intriguing candidate for
the purpose of quantum computing.

\begin{figure}
\centerline{
\epsfxsize=3.9in
\epsfbox{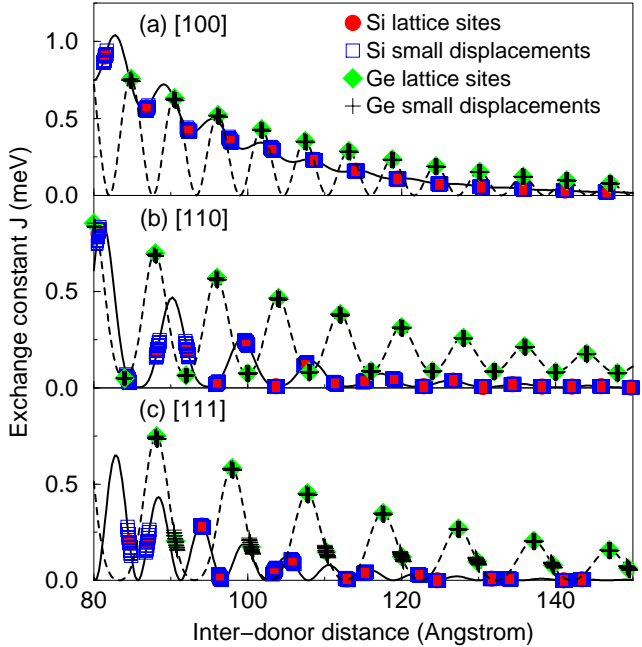}}
\vspace*{0.1in}
\protect\caption[Donor exchange in Si and Ge]{
Calculated exchange coupling between two
phosphorus donors in Si (solid lines) and Ge (dashed
lines) along high-symmetry directions for the diamond structure (see
inset in Fig.\ref{fig-hop}).
Values appropriate for impurities at substitutional sites are
given by the
circles (Si) and diamonds (Ge). Off-lattice displacements
by 10\% of the
nearest-neighbor distance lead to the perturbed values indicated
by the squares (Si) and
crosses (Ge).
}
\label{fig-JxR}
\end{figure}

We calculate the exchange energy for a pair of donors at ${\bf R}_A$ 
and ${\bf R}_B$ as a function of their relative position 
${\bf R} = {\bf R}_A - {\bf R}_B$ from Eq.(\ref{eq:exch}), taking the
expression in the square bracket equal to unity.  This means that the
results are appropriate for any lattice vector ${\bf R}$, while  
for general inter-donor distances our values should be taken as 
an estimate.  For definiteness, 
we consider ${\bf R}$ along high-symmetry
directions in the crystal lattice.  Fig.~\ref{fig-JxR} shows the
calculated values of $J(\bf R)$ for donors in Si and Ge
with $\bf R$ along
the [100], [110] and [111] directions (frames (a), (b) and 
(c) respectively). 
The solid lines give the results for Si, and exhibit  
the expected decay of $J$ with increasing $|{\bf R}|$ 
due to the decrease of the donor wavefunction overlap.  
Other general features of Si and Ge exchange are clearly
illustrated in this figure, namely the oscillatory behavior of $J$
superimposed on its overall decay with distance, and the strong
anisotropy of $J({\bf R})$ which is apparent by comparing different 
frames.
Both features are consequences of the host
material band structure and have not been considered in detail in
previous studies \cite{RKKY}, either for simplicity
\cite{Kane,Vrijen}, or because such effects are averaged out for a
random donor distribution \cite{Andres}.  
The filled circles in Fig.~\ref{fig-JxR} indicate
the accessible values of $J({\bf R})$ when the impurity pair in
Si is located at lattice sites along the considered directions.
Since for Si the conduction band minima correspond to 
points inside the BZ, the period of
oscillation in $J$ and the lattice periodicity are not commensurate.
We have also investigated the effect of small perturbations in the
atomic positions: The open squares give the resulting exchange values
when one of the impurities is slightly displaced to off-lattice
positions.  The set of squares around each circle in
Fig.~\ref{fig-JxR} corresponds to displacements 
along different
directions, with the distance from the original lattice position
arbitrarily taken as $\delta = 0.235$ \AA ~ (i.e., 10\% of the
nearest-neighbor distance in Si).  The squares follow to a very
good approximation the behavior of the calculated $J(\bf R)$, with
$\bf R$ along the unperturbed crystal direction, regardless of the
direction of $\bf \delta$. 
The upper bound of $\Delta J/J$ calculated from the
small-displacement data is about 2.5 to 5\% along the [100] direction,
which means that the requisite control over donor positioning should
be much better than 10\% of the nearest neighbor distance, i.e.
better than 0.235 $\AA$, a rather difficult task.

The corresponding results for donor pairs in Ge are given by
the dashed lines, diamonds and crosses in Fig.~\ref {fig-JxR}. 
In these calculations we have also assumed nearly free electrons,
which means $c_0^{\mu} \gg c_{{\bf K} \neq 0}^{\mu}$.
The main qualitative difference between the
calculated exchange coupling for donors in Ge and Si arises from
the different locations of the conduction band minima in the BZ. 
Since for Ge the minima occur at the zone boundary L points, the
oscillations in $J(\bf R)$ are commensurate with the lattice
periodicity.  For donors at lattice sites along the [100] direction,
the accessible values of $J({\bf R})$ correspond exactly to
successive local maxima of Eq.(\ref{eq:general}).  As a consequence,
off-lattice displacements from the original lattice sites by 10\% of
the nearest-neighbors distance in Ge ($\delta = 0.245$ \AA)
have negligible effect on the exchange coupling for donors along
this direction.  Quantitatively, the longer range for the
interactions in Ge as compared to Si is due to the larger values of
the effective Bohr radii for Ge (see Table~\ref{table1}).
\vspace*{0.2in}
Our results indicate that a $^{31}$P donor array along the [100]
direction in either of the host materials meets the requirements for
quantum computer implementation.  The exchange energy for two $^{31}$P
substitutional impurities 100 to 200 \AA apart along this direction
ranges between a few tenth of one meV to one $\mu$eV, corresponding to
less than 100 picosecond to 100 nanosecond gate operation time.  These
values are robust with respect to small off-lattice displacements.  

\vspace*{-0.2in}
\begin{figure}
\centerline{
\epsfxsize=3.8in
\epsfbox{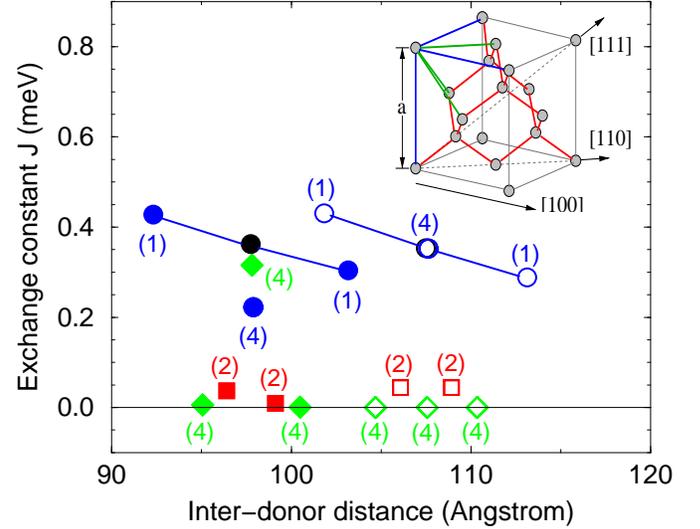}}
\vspace*{0.1in}
\protect\caption[Donor exchange in Si and Ge: effects of
hopping] 
{\sloppy{Variations in the exchange coupling between two
phosphorus donors in Si (solid symbols) and Ge (open
symbols). Lattice displacements into one of the first (red), second 
(green), or third (blue) neighboring positions
follow the color code represented in the inset, which depicts a
diamond structure with colored bonds.  The numbers in the
parenthesis next to the data points are their degeneracies,
respectively.  
For clarity, not all bonds connecting neighbors are
shown in the inset.  Black circles refer to the reference positions
${\bf R}_0$, with the two donors  
exactly along the [100] direction.  
The red squares give the exchange constant
when one of the donors is displaced into one of its nearest
neighbor sites, while green diamonds represent the 
second nearest neighbor values and blue circles 
the third nearest neighbor values.  The points connected by a 
line refer to pairs along the [100] direction, displaced by 
$\pm\rm a$ with respect to the reference position, and may be
easily identified in Fig.~\ref{fig-JxR}(a).
}}
\label{fig-hop}
\end{figure}
Displacements of the $^{31}$P atoms into
neighboring lattice sites, which are bound to occur in the
fabrication process due to either uncontrolled implantation or
surface diffusion during MBE growth, 
also deserve careful theoretical investigation.
In Fig.~\ref {fig-hop} we present the calculated $J({\bf
R})$ for a specific relative position ${\bf R}_0$ along the [100]
direction (black symbols), as well as for values of ${\bf R} = {\bf
R}_0 + {\bf  \Delta}$, with $\bf \Delta$ ranging over the 4 nearest
neighbors, 12 second nearest neighbors and 6 third nearest neighbors
in the diamond structure (see inset of Fig.~\ref {fig-hop}).
These displacements cause a relatively small change in the inter-donor 
distance (see horizontal scale in the figure). 
For third neighbors (blue symbols) the trends 
in Fig.~\ref{fig-JxR} are reasonably preserved.
Surprisingly, most first and second neighbors displacements 
essentially destroy the exchange coupling. 
This effect is entirely due to the $\cos ({\bf k}_{\mu}\cdot 
{\bf R}) $ factors in
Eq.~(\ref{eq:exch}), coming from the plane-wave part of the Bloch 
functions, 
and is not directly related to the numerical values or approximations
involved in obtaining the $j_{\mu \nu} $ coefficients. 
Therefore we do not attempt to refine the envelope function 
calculation at this stage, e.g. by including central cell corrections
\cite{Kohn} and inter-valley coupling \cite{MVE}.  The results
obtained here are inconsistent with the commonly accepted idea that
when the Bohr radii of the envelope functions are much bigger than
the lattice  parameter, the substitutional positioning of the
impurity atoms is not of much importance in determining the exchange
coupling.

In a Si QC architecture there are surface gates (so
called A and J gates) that are used to control single and two-qubit
operations \cite{Kane}.  By adjusting the gate potential(s), one can
shift the center of a donor electron wavefunction, thus quickly
traversing the fast oscillating terrain in the exchange coupling. 
The results obtained in this paper also indicate that donor electron
exchange depends very sensitively on the applied gate potential(s). 
Furthermore, whether the electrons are trapped by donor ionic
potentials or by the applied gate potentials, the fast oscillation in
exchange will persist as long as the electrons are in the bulk of
a Si crystal.  For a 2-qubit gate of a QC, it is the time
integral of the exchange constant $J$ that determines the gating
time \cite{LD}.  The implication of the oscillatory exchange is
that the A and J gate voltages corresponding to the peak exchange
coupling have to be well-controlled, optimally close to a
local maximum where the exchange is least sensitive to the gate
voltage.  Since the oscillatory exchange period is close to
lattice spacing, the positioning of the donor electrons by the A and
J surface gates must be controlled at least to that
precision.  Furthermore, if one intends to achieve adiabaticity by
increasing the switch-on time of the gate, so that $\int J dt =
(2n+1) \pi\hbar$ with  $n \gg 1$ \cite{BLD}, the error in $J$ due to
gate inaccuracies would accumulate in the integral.  It is thus
crucial to precisely control the surface gates in order to minimize
the possible errors. 

Placing the donors and therefore their electrons in the middle of
a symmetric Si quantum well will help reduce the fast
phase oscillations because only two of the six bulk Si
conduction band valleys would contribute to the donor electron ground
state \cite{Kane1}, suppressing the magnitude of the interference
effect.  In a symmetric quantum well the splitting between the ground
state and the next excited state (in analogy to the $A_1$ and $T$
donor electron states in bulk) is much smaller than in the bulk
(crudely estimated to be about 3 meV from symmetry arguments, in
contrast to nearly 12 meV in bulk Si; the smaller splitting implies
that the adiabatic condition is somewhat harder to satisfy in the
quantum well).  Strains, electric fields, and/or asymmetry in the
quantum well might help eliminate the remaining valley degeneracy
\cite{Vrijen} so that all the fast oscillating factors in the
exchange constant $J$ may be removed, thus leading to a slowly
varying $J$ and a much easier control of the exchange gate.  However,
strains themselves are hardly controllable, while asymmetry in a
quantum well may introduce additional complications that lead to
decoherence.  Further investigations of these factors are still
ongoing and the results will be reported elsewhere \cite{KHD}.

As we have demonstrated above, moving one of the donors to its 
second nearest neighbor position causes strong suppression in
exchange coupling between the two donor electrons.  This cancelation
of exchange due to valley interference might be useful for
isolating the neighboring qubits.  In a Si QC, for most of the
time the qubits should not ``talk'' to each other.  Unwanted
interaction would lead to leak of information and decoherence.
Therefore, the positions with vanishing exchange interaction might
provide a quieter environment compared to an arbitrary pair of
positions for qubits.

We thank Ravin Bhatt, Bruce Kane, Ellen Williams, Bob Clark, 
Rodrigo Capaz, and Luiz Davidovich 
for helpful discussions, and ARDA and NSA for financial support. BK 
acknowledges financial support from CNPq (Brazil).

\end{document}